\documentclass[prb,preprint]{revtex4-1} 

\usepackage{amsmath}  
\usepackage{amsfonts} 
\usepackage{graphicx} 
\usepackage{amssymb}

\begin{document}


\title{Generation, reflection and transmission of nonlinear harmonic waves by direct superposition of anharmonic dipoles}

\author{Hendradi Hardhienata}
\email{hendradi@apps.ipb.ac.id} 
\affiliation{Theoretical Physics Division, Department of Physics, IPB University, Jl. Meranti, Gedung Wing S, Kampus IPB Darmaga, Bogor 16680, Jawa Barat, Indonesia}


\date{\today}

\begin{abstract}
In this paper we describe how to derive the expressions for the higher nonlinear generation of waves, their
transmission and reflection for the case of a normal incidence
plane wave by direct superposition of dipoles oscillating anharmonically under the influence of the nonlinear polarization. We describe explicitly how the transmitted nonlinear harmonic wave for a depth beyond its coherence length can only propagate inside the material if the phase matching condition is fulfilled and show for the case of second-harmonic-generation that our calculation yields similar results with coupled-mode-theory (CMT). Furthermore, we show using dipole superposition how nonlinearity can arise in reflection due to absorption of the fundamental driving field inside the bulk.
\end{abstract}

\maketitle 

\section{Introduction} 

Since the discovery of femtosecond lasers,  nonlinear optics has become
one of the standard tools for material analysis. The numerous nonlinear optical phenomenon have greatly enhanced our knowledge regarding the interaction of light and matter as well as revolutionized our technology in optics \cite{Shen1989}.  Nonlinear optical
laser techniques are attractive because they provide a high temporal,
spatial, and spectral resolution, as well as its applicability to
all interfaces accessible by light \cite{vanHasselt1990}. 

The standard classical method to obtain expressions of transmission and reflection of the field inside a dielectric
is usually performed by applying the macroscopic Maxwell equation on the boundary between two mediums
where one will then obtain the Fresnel equation \cite{Jackson, Born1999, Pedrotti}. From the microscopic classical point of view, the dielectric consists of polarizable atoms or molecules, each of which is radiating in vacuum in response to the incident field and in response
to the fields radiated by the other atoms. The total radiated field in linear optics is then the superposition of the incident field with all radiated fields. The incident field, however, drives at the same time, due to their polarisability, the microscopic dipoles. The superposition of their radiation fields, adding the incident field too, generates the correct reflected and transmitted field. Mathematically this is expressed as an integral equation for the field either inside or outside the material.  The equivalence can be seen e.g. by comparing near field, far field, dispersion relation and Fresnel formulas for stratified layers, etc.\cite{Fearn1996,Born1999}.

\section{ Ewald-Oseen extinction theorem}
 Microscopically, nonlinear harmonics generation can be seen as radiation
from the charges that experience anharmonic oscillation due to an
external driving field. This view is actually an extension of the
idea in linear optics that an external field impinging on a dielectric
material produces dipole oscillations in the material and their superposition constitute a radiating field.
Using this approach, Ewald \cite{P.Ewald} and Oseen \cite{Oseen1915} were
able to demonstrate linear refraction and transmission phenomena inside
the dielectric microscopically, yielding an equivalent result if calculated
from the solution of the macroscopic Maxwell equation. Although the macroscopic Maxwell equations  
yield the correct solutions, they hide the physics.

\begin{figure}[htbp]
\begin{centering}
\includegraphics{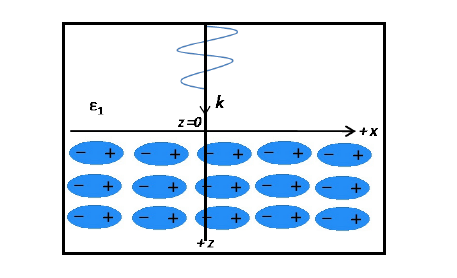}
\par\end{centering}
\caption{An incident plane wave from vacuum into a dielectric material is producing
a field polarization along the x direction. This in turn produces
dipole oscillation pararel to the polarization.}

\label{fig:figure1}
\end{figure}

 Fearn et. al. \cite{Fearn1996} has shown for the linear case that an
 incomming plane wave that is entering the material at normal incidence will produce dipole oscillations  inside the material that radiate waves at the speed of light, $c$ as shown in Fig. 1. The integral equation describing the total field at a depth $z$ from the surface is obtained by direct integration of dipoles and has the form of
\begin{equation}
E_{T}(z)=E_{i}^{\left(0\right)}e^{ik_{0}z}+\frac{ik_{0}}{2}\left[\epsilon(\omega)-1\right]\intop_{0}^{\infty}dz'E_{T}^{\left(0\right)}e^{ikz'}e^{\left|z-z'\right|}\end{equation}
In linear optics this is an integral equation, which has to be solved. For the case of nonlinear harmonic generation, however, the mathematical structure of the underlying equation to be derived will be, as seen below, much simpler. This is due to the following facts: a) There is no incoming nonlinear incident field, all the nonlinear fields are generated by the anharmonic oscillating dipoles driven by the nonlinear polarization inside the material and b) the amplitude of the nonlinear field is far smaller than the amplitude of the driving one. Strictly speaking, if the nonlinear field amplitude obtains the same size than the driving one, coupled integral equations between the linear and the nonlinear field(s) would result. However, in the low depletion approximation we assume that the amplitude of the fundamental field is space independent, and then,\emph{ instead of an integral equation, only an integration has to be performed.} Surprisingly enough, to the best of our knowledge no one has calculated this nonlinear dipole superposition explicitly even for the normal incidence case.

We proceed now in the following way. We first derive, taking the approach of Fearn~\cite{Fearn1996}, the equation for the nonlinear field. Then we derive for  a material with coherence length  $L$ and finite thicknes $z$, the superposition of the nonlinear radiation fields in transmission and compare it for the case of second-harmonic-generation (SHG) with results obtained from coupled-mode-theory (CMT) calculations. The  reflection due to integration of all nonlinear dipoles as well as the effect of field decay inside the material is also calculated.

\section{Summing Over Nonlinear Dipoles}  
Here we present a way to obtain an expression for the transmitted
and reflected nonlinear field that is produced by superposition of
all radiating (anharmonic) dipoles inside the dielectric where only the nonlinear term is treated. In doing
so, we follow a similar approach as in Ref. 5 but use
the nonlinear polarization instead of the linear case. It has to be
noted however that one should not expect to obtain the same results
as in the linear case because no second harmonic field is incidence
on the material and that the nonlinear dipoles are driven by the nonlinear fundamental external field. For simplicity, let us assume the external field
is sufficient large to generate SHG inside a dielectric and that it
is incoming at normal incidence.

To obtain the nonlinear field we calculate the contribution from all
nonlinear dipoles. We start by expressing the electric field produced
by a single dipole (Hertz dipole) at $r'$
inside the dielectric \cite{Jackson}:
\begin{equation}
\begin{array}{c}
E_{d}(r,t)=\left[3\left(\hat{p}\centerdot\hat{n}\right)\hat{n}-\hat{p}\right]\left[\frac{1}{R^{3}}p\left(t-\frac{R}{c}\right)+\frac{1}{cR^{2}}\frac{dp\left(t-\frac{R}{c}\right)}{dt}\right]+\\
\left[\left(\hat{p}\centerdot\hat{n}\right)\hat{n}-\hat{p}\right]\frac{1}{cR}\frac{d^{2}p\left(t-\frac{R}{c}\right)}{dt^{2}}\end{array}\label{eq:2-2}\end{equation}
Here $\hat{p}$ is the unit vector for the dipole polarization,    $R=\left|r-r'\right|$, and $\hat{n}$ is the unit vector directed towards $\mathbf{r}-\mathbf{r'}$. Note that as in the linear case, the retarded term $\left(t-\frac{R}{c}\right)$
 which describes the time evolution of the dipole field is propagating at the speed of light $c$,  because the atomic region is mainly vacuum. 
 
The linear case can be extended to cope with nonlinearity by considering that if the field intensity is sufficiently high, the nonlinear far fields emerges from the anharmonic motion of the charges. As in the linear case a nonlinear polarization results from this process. The nonlinear polarization for SHG (single frequency input) can be  obtained clasically by expanding
the harmonic oscillator Lorentz model assuming  nonlinear restoring forces excerted on the electron\cite{Boyd2003}:
\begin{equation}
\ddot{r}+2\gamma\dot{r}+\omega_{0}^{2}r+\beta_{2}{r}^{2}+\beta_{3}{r}^{3}+...=\frac{-e(E_{\omega}(r,t)+E_{2\omega}(r,t)+E_{3\omega}(r,t)+...)}{m}\end{equation}
where  $m, e, r$  are respectively the electron mass, charge, and its displacement from equilibrium, $-2\gamma m\dot{r}$ is the damping force and the restoring force $-\omega_{0}^{2}mr-\beta_{2}m\ddot{r}-\beta_{3}m\dddot{r}+...$ describes the nonlinear effects (second harmonic, third harmonic, ect).  Here $\beta$ is a nonlinear property of a molecule describing the second order electric suseptibility per unit volume and is called the  hyperpolarizability (tensor). Because the linear field $E_{\omega}$ is far higher than the nonlinear fields we can neglect the higher nonlinear driving fields $E_{2\omega}, E_{3\omega},$ ect. For simplicity we consider like in the linear case that the driving field comes at normal incidence into the material hence propagating along the $z$ direction.  Therefore the electric field is directed along the $x$ axis only triggering dipole oscillation also along the $x$ axis and eq. (3) becomes:
\begin{equation}
\ddot{x}+2\gamma\dot{x}+\omega_{0}^{2}x+\beta_{2}{x}^{2}+\beta_{3}{x}^{3}+...=\frac{-eE_{\omega}(z,t)}{m}\end{equation}
It has to be noted that because $E_{\omega}$ is a physical observable the value must be real. In other words $E_{\omega}$ has an explicit form of
\begin{equation}
E_{\omega}(z,t)=E_{0}\cos\left(kz-\omega t\right)=\frac{1}{2}[E_{\omega}(z)e^{-i\omega t}+ c.c. ]\end{equation}
Eq. (4) cannot be solved explicitly for $x$, but it is standard to seek a solution if the applied field $E_{\omega}(z,t)$ is considered small by means of perturbation theory. This can be done by expressing the applied field as  $E_{\omega}(z,t)=\lambda E_{\omega}(z,t)$ and writing   the displacement in terms of a power series expansion:
\begin{equation}
x=\lambda x^{(1)}+\lambda^{2}x^{(2)}+\lambda^{3}x^{(3)}+...\end{equation}
where $\lambda$ is a perturbative parameter ranging from zero to
one. The solution for $x$ and therefore the linear and nonlinear polarization $p=ex$ (where $e$ is the electron charge) can now be solved approximately and can be found in many standard textbook of nonlinear optics such as Ref. [5,6] and has the form of
\begin{equation}
p=\underset{n=1}{\sum}(\beta_{n}E_{\omega}(z')^{n}e^{-in\omega t}+c.c.)\label{eq:4}\end{equation}
 where $n$ denotes the harmonic order and $c.c.$ is the corresponding complex conjugate which ensures that the polarization has a real value. Note that $\beta$ is aside of the linear part (first term of eq. 7) a tensor.  Also it has to be stated here that we neglect local field effects which is merely adding a multiplication factor for the field. 

We next calculate the total (linear and nonlinear) dipole radiation due an incoming
field and assume for the sake of simplicity that it is incomming at normal incidence (FIG. 1). Ommiting the complex conjugate term or negative frequency for the moment in eq. (7) and insert it in eq. (2) with
$\hat{p}=\hat{x}$ and $\hat{n}\cdot\hat{x}=(x-x')/R$,
then integrating over all volumes and summing over all the terms yields for $N$ nonlinear dipoles:
\begin{equation}
\begin{array}{c}
E_{d(+\omega)}(r,t)=\underset{n}{\sum}N\beta_{n}\iiint dx'dy'dz'\left[E_{\omega}(z')\right]^{n}e^{-ik_{n\omega}R}\left[\left(3\frac{(x-x')^{2}}{R^{2}}-1\right)\left(\frac{1}{R^{3}}-\frac{ik_{n\omega}}{R^{2}}\right)-\right.\\
\left.-\left(\frac{\left(x-x'\right)^{2}}{R^{2}}-1\right)\frac{k_{n\omega}^{2}}{R}\right]\end{array}\end{equation}
Here we have used the notation $n$ to describe the higher nonlinear harmonic wave vector. It turns out that the integral in eq. (8) is easier to solve in
polar coordinates by introducing the coordinate transformation $x-x'=\rho  cos\varphi$
and $\rho^{2}=R^{2}-(z-z')^{2}$. 

The mathematical methods to perform the integration follows a similar path as in Ref. 5 with a generalized  wave vector $k_{n\omega}$ obtained in the solution instead of $k_{0}$: 
\begin{equation}E_{d(+\omega)}(z)=\underset{n=1}{\sum} n\pi ik_{n\omega} N\beta_{n}\int_{0}^{\infty}dz'\times\left[E_{\omega}(z')\right]^{n}e^{ik_{n\omega}\left|z-z'\right|}\end{equation}
The same method can be repeated  for the complex conjugate polarization term in eq. (7). One then obtains:
\begin{equation}E_{d(-\omega)}(z)=\underset{n=1}{\sum} n\pi ik_{n\omega} N\beta_{n}\int_{0}^{\infty}dz'\times\left[E_{\omega}^{*}(z')\right]^{n}e^{-ik_{n\omega}\left|z-z'\right|}\end{equation}
which differs from eq. (9) only by its minus sign. The total field is thus the sum of eq. (9) and eq. (10):
\begin{equation}E_{d}(z)=E_{d(+\omega)}(z)+E_{d(-\omega)}(z)=\underset{n=1}{\sum} n\pi ik_{n\omega} N\beta_{n}\int_{0}^{\infty}dz'\times(\left[E_{\omega}(z')\right]^{n}e^{-ik_{n\omega}\left|z-z'\right|}+c.c)\end{equation}
Eq. (11) is the total dipole contribution from the linear and nonlinear harmonics and can be used to calculate the transmission and reflection formulas for nonlinear optics. Note that it is real valued and that there is a linear propagation given by $e^{ik_{n\omega}\left|z-z'\right|}$ which is similar to the results using antenna dipole approach in calculating second harmonic generation (SHG)\cite{New2011}. 

For SHG we simply take the second term in the summation and set $n=2$ in eq. (11):
\begin{equation}
E_{d2\omega}(z)= 2\pi ik_{2\omega} N\beta_{2}\int_{0}^{\infty}dz'\times(\left[E_{\omega}(z')\right]^{2}e^{ik_{2\omega}\left|z-z'\right|}+c.c.)\label{eq:2-1-1}\end{equation}
 Before we proceed further we stop to state two important assumption
which will significantly simplify the calculation. First, because the negative frequency component that resides in the complex conjugate term in eq. (12) satisfies
\begin{equation}
\left[E_{\omega}^{*}(z')\right]^{2}e^{-ik_{2\omega}\left|z-z'\right|}=\left[E_{\omega}(z')\right]^{2}e^{ik_{2\omega}\left|z-z'\right|}
\end{equation}
we can obtain full equality in all of the following calcualtion by evaluating either the first or its complex conjugate term. Second
  for many cases, such as studying SHG in centrosymmetric media, the
nonlinear field $E_{2\omega}$  is far smaller than the linear
field $E_{\omega}$.  Therefore, the calculation can be performed independently (e.g. it does not require self consistency) 
for each nonlinear polarization term. This assumtion is similar to the low depletion approximation which allows coupled mode calculations be analyzed independently (e.g. they can be regarded as  uncoupled)\cite{Boyd2003, New2011}.

\section{ Nonlinear Transmission}
Calculation of the nonlinear transmission intensity  is usually solved either by coupled mode theory \cite{Boyd2003} or by means of the dipole antenna method\cite{New2011} which for second harmonic generation can be written in the form of\begin{equation}
I_{2\omega}\backsim\omega_{2}^{2}(E_{\omega})^{2}L Sinc^{2}\left(\Delta kL/2\right)\end{equation} where $\omega_{2}$ is the second harmonic frequency, $L$ is the penetration depth, and $\Delta k$ is the phase matching condition expressed as   $\Delta k=k_{n\omega}-nk$. As is well understood, if the total penetration depth depth of the material is smaller than the coherence length $L_{coh}=\frac{2}{\Delta k}$ a nonlinier signal is always enhanced and phase matching is not necessary as is often the case for thin films. However because generally the depth of the sample exceeds $L_{coh}$ phase matching is important to ensure constructive interference between the radiating nonlinear field and the driving field.

To calculate the nonlinear transmission for the $n$-th harmonics  we integrate (ommiting the complex conjugate because of equality) eq. (12) over z $>$ 0 and substitute as in the linear case (e.g. see eq. (32) in Ref. 5) $ \left[E_{\omega}(z')\right]^{n}=\left[E_{\omega}e^{ikz'}\right]^{n}=E_{\omega}^{n}e^{inkz'}$ where $k_{n\omega}$ is the $n$-th harmonic propagation vector.

\begin{figure}[htbp]
\begin{centering}
\includegraphics[width=9cm]{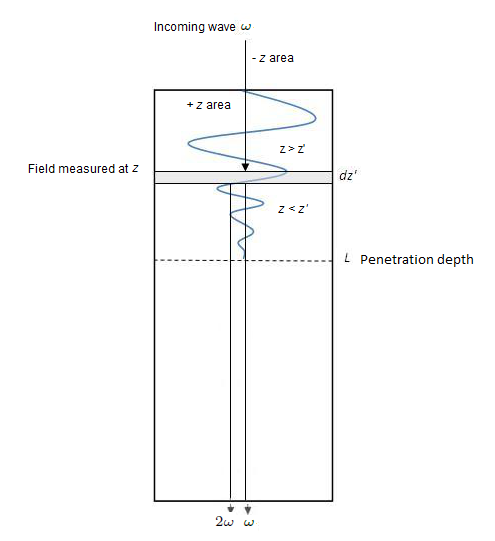}
\par\end{centering}
\caption{Coordinate system of NL dipole integration. The nonlinear far field transmitance for a normal incidence fundamental field can be calculated by summing all the radiation from the anharmonic dipoles affected over a penetration depth L}
\label{fig:figure1}
\end{figure}

 In doing the integration it is important to understand the meaning of the absolute value:
\begin{equation}
\left|z-z'\right|\left\{ \begin{array}{c}
(z-z')\:\mathrm{for}\: z>z'\\
(z'-z)\:\mathrm{for}\: z<z'\end{array}\right.\end{equation}
therefore the integral has to be splitted
\begin{eqnarray}
E_{d2\omega}(z)=  &2\pi ik_{2\omega}N\beta_{2}\left[\int_{0}^{z}dz'(E_{1})^{2}e^{i2k_{\omega}z'}e^{ik_{2\omega}(z-z')}+\int_{z}^{\infty}dz'(E_{1})^{2}e^{i2kz'}e^{ik_{2\omega}(z'-z)}\right]\label{eq:1-1}\end{eqnarray}
From Fig. 2 it is clear that for the transmission case only the integral in the first term which describe the dipole contribution from the top has to be considered from the transmission case. Thus $z$ is choosen to be the total penetration depth of the material because only over this depth a nonlinear signal is driven. Therefore we need only to integrate over $(z-z')$ from $0$ to $L$:
\begin{eqnarray}
E_{d}(z)_{n\omega} & = & 2\pi ik_{n\omega}N\beta_{n}\int_{0}^{L}dz'\times (E_{\omega})^{n}e^{inkz'}e^{ik_{n\omega}(z-z')}\nonumber \\
 & = & 2\pi ik_{n\omega}N\beta_{n} L (E_{\omega})^{n}\left[\frac{e^{i\Delta kL}-1}{i\Delta k L}\right]e^{ik_{n\omega}z}\label{eq:1-1-1-1}\end{eqnarray}
  The part in the pharanthesis denotes the evolution of the wave envelope whereas the remaining describe the phase propagation of the harmonic wave. 
Note that in  order for a nonlinear wave to propagate inside the material beyond its coherence length requires $\Delta k$   must be small or in other words the phase matching condition is necessary. It has to be noted that in doing the derivavtion we have neglected the already well understood Fresnel transmission coefficients (dependence on the refractive index) which must be considered in obtaining the correct transmission magnitude. The intensity is simply obtained by squaring the field envelope eq. (10) \begin{equation}
I_{n\omega}\backsim\omega_{n}^{n}(E_{\omega})^{n}L Sinc^{2}\left(\Delta kL/2\right)\end{equation} 
for the case of second-harmonic-generation (SHG) we set $n=2$ in eq. (10) \begin{equation}
I_{2\omega}\backsim\omega_{2}^{2}(E_{\omega})^{2}L Sinc^{2}\left(\Delta kL/2\right)\end{equation}
which is the intensity formula obtained from coupled mode theory (CMT) or the antenna picture given in eq. (14). 

\section{ Nonlinear Reflection}

The expression for reflection can be analyzed with z $<$ 0 or taking the second integration in eq. (16) and evaluate from 0 to infinity
\begin{eqnarray}
E_{d}(z)_{n\omega} & = & 2\pi ik_{n\omega}N\beta_{n}\int_{0}^{\infty}dz'\times (E_{\omega})^{n}e^{ikz'}e^{ik_{n\omega}(z'-z)}\nonumber \\
 & = & -2\pi N\beta_{n}(E_{\omega})^{n}\left(\frac{ k_{n\omega} }{nk+k_{n\omega}}\right)e^{-ik_{n\omega}z}\label{eq:1-1-3}\end{eqnarray}
Thus as in the linear case the reflection depends on $k_{\omega}$ and $k_{2\omega}$. For centrosymmetric material $\beta_{n}=0$ for all $n$ that are even numbers thus no higher harmonic generation occurs both in transmission and in reflection for SHG, FHG, ect.  Therefore the main contribution for even higher harmonic generation in centrosymmetric material must come from symmetry breaking/anisotropy in the surface where $\beta_{n}\neq 0$  or from higher order monopole contribution in the bulk (e.g. quadrupole and magnetic dipole). However an effect that is so far not yet fully understood is the role of field absorption inside the bulk which can produce a bulk dipole contribution even in a centrosymetric crystal and has so far been neglected in deriving the expression for the summation of anharmonic dipoles in transmission and reflection. Such an effect can be attributed as spatial dispersion because the field is now varying over space (in the-$z$ direction).  FIG. 3. gives an illustration to present this effect more clearly.

\begin{figure}[htbp]
\begin{centering}
\includegraphics[width=12cm]{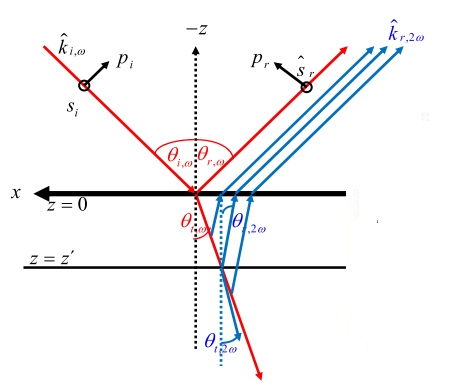}
\par\end{centering}
\caption{Bulk dipole contribution due to absorption. Besides surface and quadrupolar effects the decaying field can give rise to unequal contribution between the dipole layers along the z direction}
\label{fig:figure3}
\end{figure}

To illustrate the example more explicity we consider the SHG case and set $n=2$ in the previous calculation. When the fundamental field penetrates through the bulk, it decays with a complex wavevector $k=\omega \tilde{n} /c $. SHG field propagates with a dispersion relation given by the linear optic relation of 2$\omega$:
\begin{equation}
\varepsilon(2\omega)\cdot(2\omega)^{2}=c^{2}(k_{2\omega})^{2}\end{equation}

  To model this decay we define an absorption coefficient $\alpha_{\omega}/2$, given by $\frac{2\pi\kappa_{i}}{\lambda_{0}}=\frac{\omega\kappa_{i}}{c}$
, where $\lambda_0$ is the vacuum wavelength, and $\kappa_i$ is the imaginary part of the refractive index. The $2\omega$ wave that propagates upward also experience absorbtion by $\alpha_{2\omega}/2$. The second harmonic polarization is proportional to the the square of the fundamental field, therefore the absorption factor must be squared. Thus if we assume an exponential decaying field along the-$z$ direction the response of the dipoles along the z direction does not fully cancel each other out because each dipol in different height experience a different field. Mathematically this phenomena can be stated more explicitlty as (taking the real field):
\begin{equation}
E\left(2\omega\right)=\frac{\left(t^{1-2}\left(\omega\right)E_{0}\right)^{2}}{\chi_{r}^{(1)}\left(2\omega\right)d}N\beta_{2}k_{2\omega}\intop_{0}^{\infty}dz(\cos\left(k_{\omega}z\right)e^{-\frac{\omega}{c}\kappa_{i}(\omega)z})^{2}\cos\left(k_{\omega}z\right)e^{-\frac{2\omega}{c}\kappa_{i}(2\omega)z}\end{equation}

where we have use the real expression for the field.  Here $t^{1-2}$ is the linear Fresnel transmission coeffiecient. $N$ are the number of dipole, $\chi_{r}^{(1)}$ is the linear susecptibility and $d$ is the bulk depth. Dimensional analisys yields consistency between the right and left side of the equation and an order of magnitude estimate yields a second harmonic field of 100 $V/cm$ which is a reasonable number. Performing the integration we obtain 
\begin{equation}
\begin{split}
E\left(2\omega\right) & =\frac{\left(t^{1-2}\left(\omega\right)E_{0}\right)^{2}}{\chi_{r}^{(1)}\left(2\omega\right)d}N\beta_{2}k_{2\omega}\\
& \frac{k_{SHG}}{2}\left(\frac{2}{4k_{2\omega}+\kappa_{SHG}^{2}}+\frac{1}{4(k_{2\omega}-2k_{\omega})+\kappa_{SHG}^{2}}+\frac{1}{4(k_{2\omega}+2k_{\omega})+\kappa_{SHG}^{2}}\right)
\end{split}
\end{equation}

here we have use $\kappa_{SHG}=\frac{2\omega\kappa_{\omega}}{c}+\frac{2\omega\kappa_{2\omega}}{c}$. Evaluating the terms and assuming phase matching condition. The only term that is significant compared to the other is:

\begin{equation}
E\left(2\omega\right)\approx\frac{\left(t^{1-2}\left(\omega\right)E_{0}\right)^{2}}{\chi_{r}^{(1)}\left(2\omega\right)d}N\beta_{2}k_{2\omega}\left(\frac{k_{SHG}}{2}\frac{1}{4(k_{2\omega}-2k_{\omega})+\kappa_{SHG}^{2}}\right)\end{equation}
Note that we now got an additional absorption term in the denominator. Fig. 4 depicts the SHG reflection intensity as a function of the absorption coefficient. 

\begin{figure} [htbp]
\begin{center}
\includegraphics[width=12cm]{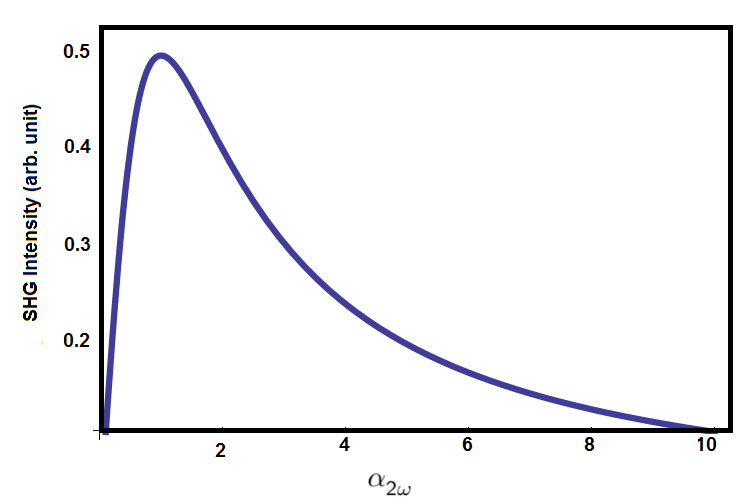}
\end{center}
  \caption{SHG intensity dependence on the nonlinear absorption coeficient}.
  \label{fig:figure3}
\end{figure}

It can be seen that if the nonlinear absorption coefficient is very small the SHG intensity is also very small. This coresponds to a material that is very transparent so the energy conversion into nonlinear intensity is very small hence a low signal is obtained in reflection. For a very large absorption the intensity is also decreasing. This is related to very small light penetration into the bulk thus only a few dipoles inside of the material can radiate SHG. 

\section{Summary}
We show by direct superposition of dipoles that the formulas of reflection and transmission for the nonlinear harmonics can be obtained. The results also show that phase matching is an important requirement for nonlinear wave propagation and that dipole contribution can arise from the bulk of a centrosymetric material aside from surface and quadrupole effects in reflection  due to absorption of the driving field.

\emph{Acknowledegements}: The author would like to thank Kurt Hingerl, Adalberto Alejo-Molina, and Norbert
Esser for stimulating discussions and acknowledges research funding from the Hibah Kompetensi Grant  No. 3989/IT3.L1/PN/2020 ).

\bibliography{SHGSiReferencesEwald}

\end{document}